\documentclass[12pt]{article}

\usepackage{amsmath,amstext,amsfonts}

\newcommand{\mvec}{{\bf m}}
\newcommand{\nvec}{{\bf n}}
\newcommand{\kvec}{{\bf k}}

\newcommand{\mA}{{\bf A}}

\newtheorem{theorem}{Theorem}[section]

\newtheorem{lemma}[theorem]{Lemma}

\def\bd{\begin{displaymath}}
\def\ed{\end{displaymath}}

\def\eqref#1{(\ref{#1})} 
\def\qed{\hbox{\hskip 6pt\vrule width6pt height7pt depth1pt
    \hskip1pt}\bigskip}

\def\runinend{\enspace}
\def\ackname{Acknowledgement\runinend}%
\def\acknowledgements{\par\addvspace{17pt}\rmfamily
\def\ackname{Acknowledgements\runinend}%
\trivlist\if!\ackname!\item[]\else
\item[\hskip\labelsep
{\bf\ackname}]\fi}%

\begin{document} 
\bibliographystyle{plain} 

\hfill{} 
 
\thispagestyle{empty}

\begin{center}{\bf Persistent energy flow for a  
stochastic wave equation model in nonequilibrium statistical 
mechanics } 
\vspace{5mm}

{\bf Lawrence E. Thomas}  \\    

{\small\it Department of Mathematics, University of Virginia, \\
Charlottesville, VA 22904 \\ }

\end{center}

\setcounter{page}{1}

\begin{center} {\it Dedicated to Elliott Lieb on the occasion of his $80^{th}$ birthday.} \end{center}

\begin{abstract} We consider a one-dimensional partial differential equation system
  modeling heat flow around a ring. The system includes a Klein-Gordon
  wave equation for a field satisfying spatial periodic boundary
  conditions, as well as Ornstein-Uhlenbeck stochastic differential equations
  with finite rank dissipation and stochastic driving terms modeling
   heat baths.
  
      There is an energy flow around the ring. In the case 
of a linear field with different (fixed) bath temperatures, the energy flow
can persist even when the interaction with the baths is turned off.  A simple
example is given.        
 \end{abstract}
 

\noindent Running head: 
Persistent energy flow for a stochastic  wave equation\\


\section{Introduction} We consider the following system 
of coupled partial differential equations:
\begin{eqnarray}
\partial_t\phi(x,t)&=& \pi(x,t)\nonumber\\  
\partial_t\pi(x,t)&=& (\partial_x^2 - 1)\phi(x,t)- g(\phi(x,t))-\eta r(t)\cdot\alpha(x)
\nonumber\\
dr_i(t)&=&  -\left(r_i (t)-\eta\langle\alpha_i,\pi(t)\rangle\right)dt + 
\sqrt{T_i}d\omega_i(t)\,\,i= 1,2. \label{theeq}
\end{eqnarray}

In these equations $\phi$ is a field and $\pi$ is the corresponding momentum
field;  both satisfy periodic boundary conditions with $x \in
[0,2\pi]$. The non-linear term $g$ is assumed to be bounded Lipschitz.
The functions (actually distributions) $\alpha =
(\alpha_{1},\alpha_{2})$ are fixed, real, and assumed to have Fourier
coefficients $\hat{\alpha_i}(n) \propto n^{\theta}$ for some
$-1/2<\theta<1/4$, made precise below. (The brackets
$\langle\cdot,\cdot\rangle$ stand for the $L^2[0,2\pi]$-inner product or
distributional-test function pairing.)  The vector $r=(r_{1},r_2)\in
{\mathbb R}^2$ is an artifact of the baths, and $\eta$ is a
coupling constant controlling the strength of the coupling of the
baths to the field.  Finally, $\omega = (\omega_{1},\omega_2)$ is
$2$-dimensional standard Brownian motion,  and $T_1$ and $T_2$ are the bath
temperatures. The system of equations \eqref{theeq} serves as a model of
 heat conduction around the ring.

    We will show, for the linear setting with $g\equiv 0$
 and with the system in a stationary non-equilibrium state,  i.e. with the
 temperatures $T_1$ and $T_2$ {\it differing},  that  there
 can be a stationary energy flow--a current--around the ring and, moreover,
 that  this current persists even as the bath couplings are turned off, $\eta\rightarrow 0$. This phenomenon was
 noted but not proved in a previous article \cite{TW1} with Y. Wang.
The goals of the present article are to provide a proof of the claim and to give
 an explicit  formula for the current. Parts of the analysis can be
 carried out assuming a non-linearity $g$ in the wave equation, and so
  the non-linearity is retained in the analysis to the extent
  possible,  with a view toward understanding its effect on
 this persistent current.
 
The model itself has its origins in the heat flow models studied by
Eckmann, Pillet, and Rey-Bellet \cite{EPR1,EPR2}, who considered a
finite chain of non-linear oscillators coupled at either end to
free fields that model heat baths. Assuming a judicious choice of couplings and Gaussian-distributed random 
initial conditions on the baths, these authors showed
that, after
integrating out the bath field equations, one is left with Markovian
stochastic equations. They established existence of a unique
stationary state exhibiting heat flow and entropy production for these equations.  
See also \cite{EH1,EH2,EH3, RT1,RT2,RT3} for
further developments of the model.  The equations above,
\eqref{theeq}, are an adaptation of the models to an
infinite-dimensional setting where the chain of oscillators has been
replaced by a wave equation.

Even in the case where $g$ is cubic, $g(y)= y^3$ and thus unbounded, and the
$\alpha$'s are chosen appropriately, the equations of motion \eqref{theeq} are known to have global in time
solutions  in spaces of low regularity, that is, with  $\phi(\cdot,t)$  in
$H^{\gamma}$, for any $\gamma>1/3$ ($\gamma=1$ corresponding to the usual energy
norm) \cite{RT4}.
 For the linear problem $g= 0$, the equations are
of Ornstein-Uhlenbeck form. They have a unique invariant Gaussian
measure (regardless of whether the temperatures are equal or not),
and this measure is supported on fields $(\phi,\pi)$ such that
$\phi(x)$ is almost surely H\"{o}lder continuous with index $<1/2$ \cite{TW1}.    
For bounded, Lipschitz $g$, the equations of motion with ultra-violet cutoffs
have a unique invariant measure for each cutoff.  By a soft argument,
these measures have a weak* limit as the cutoff is removed, but it is
not known whether the limiting measure(s) is invariant with respect
to the above equations except in the equilibrium case $T_1= T_2$,
where a Gibbs state for the field is readily seen to be invariant 
\cite{Mc,Zh,RT4,TW2}.

We turn now to the principal result. 
 Let $\hat{\alpha}_i(n)$ be the $n^{th}$ Fourier
coefficient of $\alpha_i$,
\begin{equation}
        \hat{\alpha}_i(n)\equiv
        \int_0^{2\pi}\!\!\!\!e^{-inx}\alpha_i(x)\,dx.
\end{equation}

\noindent {\bf Assumptions}.  {\it We assume that there exist positive constants $c_1$
and $c_2$ and another constant $\theta$, $-1/2<\theta< 1/4$ such that
for
all $n\neq 0$, 
\begin{equation}\label{assumption1.eq}
            c_1 |n|^{\theta}\leq |\hat{\alpha}_i(n)|<c_2
            |n|^{\theta}\,\,i= 1,2
\end{equation}
and
\begin{equation}\label{assumption2.eq}
         c_1(|\hat{\alpha}_1(n)|^2+|\hat{\alpha}_2(n)|^2)\leq |\hat{\alpha}_1^2(n)+\hat{\alpha}_2^2(n)|^2. 
\end{equation}}

By way of explanation, the equations of motion
are such that high frequency Fourier modes of the field couple rather
weakly to the noise; however, the dissipation of these modes also is 
weak.  Eq.(\ref{assumption1.eq}) insures that
the modes couple sufficiently strongly to the dissipation  to
counterbalance the driving terms, thereby maintaining stationarity.
(The situation is to be contrasted with stochastic heat equations and
equations of hydrodynamics with viscosity, where high frequency modes
dissipate very rapidly \cite{BKL,EMS,FM,HM,KS}.)  The second condition,
Eq.\eqref{assumption2.eq}, insures that degeneracy of mode frequencies is
broken with the coupling $\alpha$'s turned on. As will be seen, the difference
of frequencies between modes is the primary factor in controlling the
correlations between field modes.

  The classical energy flow through  $x\in [0,2\pi]$ is given
by $\pi(x)\partial_x\phi(x)$. We will consider this energy flow
averaged over the ring, i.e., $\frac{1}{2\pi}\int \pi(x) \partial_x
\phi(x)\,dx$.  A non-zero expectation of this quantity in a stationary
state signals a net circular current.  We 
project the equations of motion Eq.(\ref{theeq}) onto the subspace spanned by $\{e^{inx}\}$,
$|n|\leq M$, and we let $E_M$ denote expectation with respect to
the invariant measure for the system in this subspace. Although it is in fact a long
calculation, we state the principal result as a
theorem:

\begin{theorem} For the linear system $g\equiv 0$, we have 
\begin{eqnarray}\label{Maintheorem.eq.1}
\lefteqn{\lim_{\eta\rightarrow0} \lim_{M\rightarrow\infty}E_M\left[\frac{1}{2\pi}\int_0^{2\pi}\!\!\!\!\!\!\pi(x)\partial_x\phi(x)\,dx\right]}&&\\
&=& -\frac{(T_1-T_2)}{2}\frac{1}{\pi}\sum_{n:\,n>0} \frac{n \Im\left(\hat{\alpha}_1^{*\,2}(n) \hat{\alpha}_2^2(n) \right)}
 {\left((n^2+1)|\hat{\alpha}_1^2(n) +\hat{\alpha}_2^2(n)|^2 +(|\hat{\alpha}_1^2(n)| +|\hat{\alpha}_2^2(n)|)^2\right)}.\nonumber
\end{eqnarray}
\end{theorem}

The theorem requires explanation.  First, we emphasize that the
result concerns only the linear problem $g=0$. Second, for the equilibrium state or for a
non-equilibrium limiting stationary state, $\pi(x)$ is in $H^{\gamma-1}$,
$\phi(x)$ in $H^{\gamma}$, with $\gamma<1/2$ almost surely, so that
the pointwise heat flow $\pi(x)\partial_x\phi(x)$ makes no sense; however, the
product does make sense for the equations with an ultraviolet
cutoff $M$. There is persistent current flow for the
finite-dimensional ultraviolet cutoff systems.  The principal result here is the {\it uniform}  convergence
of this current, ${M\rightarrow \infty}$ in $\eta\rightarrow 0$, and 
indeed that the current {\it has} a limit. It is this uniformity that we
investigate here in detail, while including the effect of non-linearity
where possible. Finally, we remark that  when the coupling functions are turned
off, the linear equations of motion certainly are not ergodic, and indeed
invariant measures other than GIbbs states are possible. 
Of particular interest is how different non-equilibrium states are singled out in this
limit, $\eta\rightarrow 0$.

The existence of such a current was announced
in joint work with Y. Wang \cite{TW1}.   What is new here is an
account of the analysis behind the computation of the current, as well as a more
explicit form for it in terms of the Fourier series coefficients of
the coupling functions.  We also provide an example  below in which the
coupling $\alpha$'s are $\delta$-functions at two points $x_0,
\,x_1\in [0,2\pi)$-- pictorially the situation in which the heat
reservoirs are attached to a vibrating ring at just two points. The persistent current in this case
 shows a surprising discontinuous behavior
as a function of the separation $x_1-x_0$ of the two contact points.

The result here, for a linear system, is  in the spirit of the 1966
paper by Rieder, Lieb, and Lebowitz \cite{LLR} on stationary states for
harmonic crystals.  These authors considered a stationary state for a
linear chain of oscillators with thermal reservoirs at either end
maintained in general at different temperatures.  They found that the
expected kinetic energy of the $j^{th}$ oscillator is essentially
constant in $j$ in the interior of the chain (at the mean
temperature), but that it {\it dips} as a function of $j$ near the
hotter end, with then a large discontinuous positive spike at the end
oscillator.  Similiarly, the expected kinetic energy {\it grows} in $j$ in a
neighborhood of the cooler end with a large discontinuous dip at the end
oscillator.  They also determined the expected energy flow, 
which is constant throughout the chain but which  exhibits  anomalous  behavior 
when the coupling is large.  Presumably these effects disappear with non-linearity,  as might well
happen to the current in our model if the non-linearity effectively  impedes energy flow around the ring.

We remind the reader that the existence of a stationary state invariant under the evolution equations 
for the non-linear ($g\neq 0$), nonequilibrium system remains 
an open question. One might attempt to construct an ultraviolet 
limit of stationary measures for $g\neq 0$, $M\rightarrow\infty$, in such a
way that the expected current is fixed and finite.  This would suggest 
how $g$ is to be renormalized (if indeed renormalization is necessary) in order to achieve this limit. 

\subsection{Example}

We conclude this introduction with a simple example in which the coupling distributions, the
$\alpha$'s, 
are chosen to be $\delta$-functions, in particular
\begin{equation}
        \alpha_1(x)= \delta(x),\,\,\alpha_2(x)= c\delta_2(x-x_1)
\end{equation}
with $x_1$ chosen $0<x_1\leq \pi$, and $c$ chosen $0< c <1$.  In this
case $\hat{\alpha}_1(n)= 1$ and $\hat{\alpha}_2(n)= c\,
e^{-inx_1}$, and the growth (with $\theta=0$) and non-degeneracy
conditions on the $\alpha$'s, Eqs.(\ref{assumption1.eq},\ref{assumption2.eq}),
 are seen to be satisfied.  
The limiting current is then given by
\begin{equation}
C(x_1)\equiv  \frac{T_1-T_2}{2\pi} \sum_{n: n\geq 1} \frac{n c^2 \sin(2n x_1)}{(n^2+1)(1+c^4 +2c^2\cos(2nx_1)) +(1+c^2)^2}\,,
\end{equation}
which is a conditionally convergent series.  
When the $\delta$-functions are localized
at diametrically opposite points, i.e., at $0$ and at $x_1= \pi$,  this
expected limiting current  $C(x_1)$ is zero,  
as one  expects  just for symmetry reasons. It is perhaps strange that $C(x_1)$
also vanishes for $x_1= \pi/2$.  As seen numerically,  $C(x_1)$ exhibits discontinuous jumps at both $\pi/2$ and $\pi$
 but  is otherwise continuous in $x_1$. We have
no physical explanation for this behavior.

\section{Calculation of the expected current}

\subsection{Notation and eigenfunctions} 

Let
   \begin{equation}
               \mA(\eta) = \left( \begin{array}{cccc}
               0&1&0&0\\
               \partial_x^2-1& 0&-\eta \alpha_1&-\eta \alpha_2 \\
               0&\eta \langle \alpha_1 &-1& 0\\
               0&\eta \langle \alpha_2 &0 &-1
               \end{array} \right).
\end{equation}
In terms of $\mA(\eta)$, the linear equations of motion Eq.(\ref{theeq}) read 
               \begin{equation}
                                    d\Phi(t)= \mA(\eta)\Phi(t) +\sqrt{T} d\omega,
                                    \end{equation}
with $\Phi(t)$ the four-vector $\Phi(t)= (\phi(x,t),\pi(x,t),r_1(t),r_2(t))^t$.  

We will utilize the eigenfunctions and eigenvalues of $\mA(\eta)$ obtained via perturbation theory,
$\mA(\eta)$ being  a rank 2 perturbation of $\mA(0)$.  (See particularly the appendix
of \cite{TW2} for an account of the estimation of these eigenfunctions and their eigenvalues.)  Note that $\mA(\eta)$ is not normal for $\eta\neq 0$.
We denote the right (column) eigenvectors of $\mA(\eta)$
by $e_{\nvec}(\eta)= (e_{\nvec\phi}(\eta), e_{\nvec\pi}(\eta), e_{\nvec r}(\eta))^{t}$, the last component being an abbreviation for the two $r$ components. The subscript $\nvec$ 
is a triple, $\nvec= (\pm, n,\sigma)$, $n\geq -1$ being an integer, $\sigma = 1,2$. The eigenvalue $\lambda_{\nvec}(\eta)$ corresponding to $e_{\nvec}(\eta)$ is equal to
           \begin{eqnarray}\label{eigenshift5appendix.eqn10}
\lambda_{\nvec}(\eta)=  {\lambda_{\pm,n,\sigma}}(\eta)
&=& \pm i(n^2+1)^{1/2} -\frac{\eta^2\mu_{n,\sigma}}{2(\pm i(n^2+1)^{1/2} +1)}\nonumber\\
&&\phantom{XX} +{\cal O}(\eta^4 n^{4\theta-2}\ln{n})
 \nonumber\\
 \end{eqnarray} 
% +{\cal O}(\eta^4 n^{4\theta-3}\ln{n})
  for $n$ large (where by abuse of notation, ${\cal O }(\eta n^\theta)= {\cal O}(\eta){\cal O}(n^\theta)$, for example).
The eigenvalue shifts $\mu_{n,\sigma}$, $\sigma = 1,2$ are the eigenvalues of the $2\times 2$-matrix
\begin{equation}\label{twobytwo.eq}
   M(n)\equiv\left(\begin{array}{cc}
         |\hat{\alpha}_1^2(n)|+|\hat{\alpha}_2^2(n)|& \hat{\alpha}_1^2(n)+
         \hat{\alpha}_2^2(n)\\
 \hat{\alpha}_1^{*\,2}(n)+  \hat{\alpha}_2^{*\,2}(n) & |\hat{\alpha}_1^2(n)| + |\hat{\alpha}_2^2(n)|
   \end{array}\right)
  \end{equation}
and are given by
\begin{equation}\label{mueigen.eq}
        \mu_{n,\sigma}= |\hat{\alpha}_1(n)|^2 + |\hat{\alpha}_2(n)|^2 -(-1)^{\sigma}\left|\hat{\alpha}_1(n)^2 +\hat{\alpha}_2(n)^2\right| , \,\,\sigma= 1,2.
\end{equation}
For $n= -1$, the two corresponding eigenvalues of $\mA(\eta)$ are simply $\lambda_{-1,\sigma} = -1 +{\cal O}(\eta^2)$; the $\pm$ plays no role.  For the unperturbed
$\eta=0$ eigenfunctions, we take, for $n> 0$, 
             \begin{eqnarray}
                      e_{\pm, n, 1}(0)&= &\frac{1}{2\sqrt{\pi}}(\mp i (n^2+1)^{-1/2}\cos (nx +\psi_n), \cos(nx+\psi), 0)^t \nonumber\\
                      e_{\pm, n, 2}(0)&=& \frac{1}{2\sqrt{\pi}}(\mp i (n^2+1)^{-1/2}\sin(nx +\psi_n), \sin(nx+\psi), 0)^t,
              \end{eqnarray}    
where $\psi_n$ is the phase defined by
\begin{equation}
               e^{i\psi_n}= \frac{\hat{\alpha}_1^2(n) +\hat{\alpha}_1^2(n) }{\left|\hat{\alpha}_1^2(n) +\hat{\alpha}_2^2(n)\right | }.
\end{equation}
For $n=0$, $e_{\pm,0}= \frac{1}{2\sqrt{\pi}}(\mp i, 1, 0)^t$; these eigenfunctions are not degenerate and there is no $\sigma$ index. For $n= -1$, $e_{-1,1}= (0,0,1,0)$ and
$e_{-1,2}= (0,0,0,1)$, the $\pm$ index is not in play. 
The momentum components of these functions, projected into the subspace of $L^2(0,2\pi)$ spanned by $e^{\pm i n x}$, are eigenvectors of
$M(n)$ above. 

 Similarly, we have left eigenvectors (linear functionals) $\{f_{\nvec}(\eta)\}$ of $\mA(\eta)$,  their unperturbed versions taking the form
             \begin{eqnarray}
                      f_{\pm, n, 1}(0)&= &\frac{1}{2\sqrt{\pi}}(\pm i (n^2+1)^{1/2}\cos (nx +\psi_n), \cos(nx+\psi), 0)^t \nonumber\\
                      f_{\pm, n, 2}(0)&=& \frac{1}{2\sqrt{\pi}}(\pm i (n^2+1)^{1/2}\sin(nx +\psi_n), \sin(nx+\psi), 0)^t,
                                   \end{eqnarray}      
with analogous expressions for $f_{\pm,0}$ and $f_{-1,\sigma}$.  Both the $e_{\nvec}(\eta)$'s  and the $f_{\nvec}(\eta)$'s are normalized so that
 the real inner product $\langle f_\nvec(\eta), e_{\nvec}(\eta) \rangle = 1$
and so that, for $n\geq 0$, $e_{\nvec,\pi}(\eta) $ has $L^2(0,2\pi)$-norm equal to $1/2$.   
We emphasize that we are particularly concerned with the large $n$ behavior of these functions. Also, here and throughout, we are regarding
these $f_{\nvec}$'s as linear functionals so that, e.g., 
\begin{equation}
                 \Phi(f_{\nvec}(\eta))\equiv \int_0^{2\pi} \!\!\!\big(f_{\nvec,\phi}(\eta, x)\phi(x) +f_{\nvec,\pi}(\eta, x)\pi(x)\big) dx + f_{\nvec,r}(\eta)\cdot r.
                 \end{equation}

We have that the momentum components of $e_{\nvec}(\eta)$ and $e^o_{\nvec}(0)$
differ by a small amount:   
\begin{lemma}{\rm (Cf. \cite{TW2})} For $\nvec= (\pm, n,\sigma)$
\begin{equation}\label{ee.eqn}
         e_{\nvec,\pi}(\eta)-e_{\nvec,\pi}(0)= \sum_{\kvec}a_{\nvec,\kvec}(\eta) e_{\kvec,\pi} (0)
\end{equation}
where the Fourier coefficients $ a_{\nvec,
  \kvec}(\eta)$ are estimated
\begin{eqnarray}\label{fouriercoefficients.eq}
      a_{\nvec, \kvec}(\eta)&=& \left\{\begin{array}{ll}
                {\cal O}(\eta^4 n^{4\theta-2}\ln^2 n)     &\kvec = \nvec \,\,{\rm or} \,\,\kvec= (\mp,n,\sigma)\\ 
                {\cal O}(\eta^2 n^{2\theta-1}\ln n)     &\kvec =
                (\pm,n,\sigma'),\,\,{\rm or} \,\,\kvec= (\mp,n,\sigma'),\,\,\sigma'\neq \sigma\\
               {\cal O}(\eta^2\frac{k^{\theta}n^{\theta}}{n^2-k^2})     &\kvec= (\pm,k,\sigma'), {\rm with}\,\, 
                  k\neq n.
                \end{array}  
               \right.
\end{eqnarray}
Moreover, we also have that the imaginary parts of these coefficients 
satisfy 
\begin{eqnarray}\label{fouriercoefficientsim.eq}
      \Im\, a_{\nvec,\kvec}(\eta)&=& \left\{\begin{array}{ll}
                {\cal O}(\eta^4 n^{4\theta-3}\ln^2 n)     &\kvec = \nvec\,\,{\rm or} \,\,\kvec= (\mp,n,\sigma)\\ 
                {\cal O}(\eta^2n^{2\theta-2}\ln n)     &\kvec =
                (\pm,n,\sigma')\,\,{\rm or} \,\,\kvec= (\mp,n,\sigma'),\,\,\sigma'\neq \sigma\\
                {\cal O}\left(\frac{\eta^2 k^{\theta}n^{\theta-1}}{n^2-k^2}\right)     &\kvec= (\pm,k,\sigma'), {\rm with}\,\, 
                  k\neq n.
                \end{array}  
               \right.
\end{eqnarray}
These estimates are uniform  with respect to an ultraviolet cutoff $M$.
\end{lemma}

We refer particularly to the appendix of \cite{TW2} for details of these computations; although  tedious, they just involve finite rank degenerate perturbation theory.  The basic idea
is to write the eigenvalue equations as two-dimensional implicit eigenvalue equations. We add:\\

\noindent {\it Remark}. In these estimates and throughout this article one  encounters sums  of the sort $\sum_{k >0:\,k\neq n}\frac{k^{\theta}}{|k^2-n^2|}$. A convenient
method for estimating these sums is to break the sum up into three pieces: 1) an ``infrared" part with $k<n/2$ and where the denominator is simply replaced by $n^2/2$, the $k$-sum
then giving ${\cal O}(n^{\theta+1}/ n^2)$;  2) a ``near" part where $n/2\leq k< 3n/2$,  where the numerator is treated as a constant $3n^{\theta}/2$ and the denominator
is replaced by ${n |{n-k}|}$ with the resulting $\sum_{k: \,n/2\leq k< 3n/2} \frac{1}{|n-k|}= {\cal O}(\ln n)$, so that the near part then gives ${\cal O}(n^{\theta -1}\ln n)$;  and 3) an ``ultraviolet" part with $k\geq 3n/2$, where the summand is estimated simply by $n^{\theta-2}$
with resulting sum ${\cal O}(n^{\theta -1})$.  Thus this example would give,  overall, ${\cal O}(n^{\theta-1}\ln n)$, furnishing some 
explanation for the mysterious logarithms in our estimates.

\subsection{Averaged expected energy flow}

In this section, we give an expression for the  expectation with respect to the invariant measure of the
energy flow $\pi(x)\partial_x\phi(x)$ through a point $x\in [0,2\pi]$ averaged over the circle.  We assume an
ultraviolet cut-off $M$; Fourier modes of the field with frequency $|n|>M$ are simply set to zero, but all of our estimates are uniform with respect to the cutoffs, so 
the cutoff $M$ is not explicitly expressed. 
Thus we consider
\begin{equation}\label{doublesum.eqn1}
  E\left[\frac{1}{2\pi}\int_0^{2\pi}\!\!\!\!\!\pi(x)\partial_x\phi(x) dx\right]=
 \frac{1}{2\pi} \sum_{\mvec}\sum_{\nvec}\langle e_{\mvec,\pi}(\eta),\partial_x e_{\nvec,\phi}(\eta)\rangle
  E[\Phi(f_{\mvec}(\eta))^*\Phi(f_{\nvec}(\eta))],
\end{equation} 
where on the right side we have employed eigenfunction expansions for $\pi(x)$ and $\phi(x)$ using the eigenfunctions
of $\mA(\eta)$.  The inner product is simply the $L^2(0,2\pi)$-inner product involving the $\pi$ and $\phi$ components of the eigenfunctions.  In order to prove the
theorem, we will show that the diagonal terms, i.e. terms with $\mvec = \nvec$ in this double sum, do not contribute in the limit $\eta\rightarrow 0$, nor do terms $\mvec, \nvec$ with
$m\neq n$. Only the near-resonant terms, terms such as $\mvec= (\pm,n,1)$ and $\nvec= (\pm,n,2)$, contribute in the limit, taking the form stated in the theorem.
  \\

\noindent {\it Diagonal terms}\\

We begin by considering the diagonal terms in the above equation (\ref{doublesum.eqn1}), terms with
$\nvec= \mvec$.        
The estimates  Eqs.(\ref{fouriercoefficients.eq},\ref{fouriercoefficientsim.eq}) above and the fact that $e_{\nvec,\phi}(\eta)= \frac{1}{\lambda_{\nvec}(\eta)} e_{\nvec,\pi}(\eta)$ from its
eigenvalue equation
imply that
\begin{eqnarray}\label{matrixdiag.eq}
    \langle e_{\nvec,\pi}(\eta),\partial_xe_{\nvec,\phi}(\eta)\rangle&=&
   \frac{2i}{\lambda_{\nvec}}\langle
   \Re\,e_{\nvec,\pi}(\eta),\partial_x\Im\,e_{\nvec,\pi}(\eta)\rangle\nonumber\\
&=& {\cal O}(\eta^2 n^{2\theta-2}\ln n) +{\cal O}(\eta^4 n^{4\theta-2}).
\end{eqnarray}
Now in the linear theory, $ E[|\Phi(f_{\nvec})|^2]$ is bounded by the supremum of the temperatures, uniformly  in $\eta$ small,
see \cite{TW1,TW2},  so that  $ \langle e_{\nvec,\pi}(\eta),\partial_xe_{\nvec,\phi}(\eta)\rangle E[|\Phi(f_{\nvec})|^2]$ is certainly summable
over $\nvec$ provided  the  exponent $\theta< 1/4$ as is assumed;  the sum clearly vanishes in the limit $\eta\rightarrow 0$.  (In a non-linear theory, $g\neq 0$, we 
would also need  $E[|\Phi(f_{\nvec})|^2]$ to be bounded or have suitably slow growth in $n$, but this is not known except in equilibrium,
$T_1= T_2$.) \\

\noindent{\it Off-diagonal terms}\\

We next consider the non-resonant off-diagonal terms $\mvec\neq \nvec$ with $m\neq n$ or the 
$\pm$'s differing in the double sum Eq.(\ref{doublesum.eqn1}).  The above estimates of the lemma,
Eqs.(\ref{ee.eqn}, \ref{fouriercoefficients.eq}), imply that 
\begin{eqnarray}\label{matrixfar.eq}
      \langle e_{\mvec,\pi}(\eta),\partial_x e_{\nvec,\phi}(\eta)\rangle&=& {\cal
      O}\left(\eta^2 \frac{m^{\theta}n^{\theta}}{m^2-n^2}\right)+{\cal
      O}\left(\eta^2\frac{m^{\theta+1}n^{\theta-1}}{m^2-n^2}\right)\\
      &&\!\!\!\!\! \!\!\!\!\! +  {\cal O}(\eta^4 m^{\theta}n^{\theta-1})\sum_{k:\,k\neq m,n} {\cal O}\left(  \frac{k^{2\theta+1} }{(k^2-m^2)(k^2-n^2)}\right) ,\,\,m\neq n.\nonumber
\end{eqnarray}    
 The last term on the right side of this equation can be estimated using the Schwarz inequality (regarding
 one of the functions as $k^{\theta}/ (k^2-m^2)$):
 \begin{equation}\label{OO.eq}
       {\cal O}( \eta^4m^{\theta}n^{\theta-1})\sum_{k:\,k\neq m,n} {\cal O}\left(  \frac{k^{2\theta+1} }{(k^2-m^2)(k^2-n^2)}\right)
       = {\cal O}(\eta^4m^{2\theta -1} n^{2\theta-1}) .   
\end{equation}
Thus
\begin{eqnarray}\label{matrixfar2.eq}
      \langle e_{\mvec,\pi}(\eta),\partial_x e_{\nvec,\phi}(\eta)\rangle&=& {\cal
      O}\left(\eta^2\frac{m^{\theta}n^{\theta}}{m^2-n^2}\right)+{\cal
      O}\left(\eta^2\frac{m^{\theta+1}n^{\theta-1}}{m^2-n^2}\right)\nonumber\\
      && + {\cal O}( \eta^4m^{2\theta -1} n^{2\theta-1}).    
\end{eqnarray}    
    
We also need an estimate on the correlation $ E[\Phi(f_{\mvec}(\eta))^*\Phi(f_{\nvec}(\eta)]$. Simply differentiating this
    expression with respect to time and using stationarity, we obtain the identity
    \begin{eqnarray}\label{stationident.eq}
      \lefteqn{E[\Phi(f_{\mvec}(\eta))^*\Phi(f_{\nvec}(\eta))]}\nonumber\\
       &=& \frac{1}{\lambda^*_{\mvec}(\eta)+\lambda_{\nvec}(\eta)}\Big(E[ \Phi(f_{\mvec}(\eta))^* \langle f_{\nvec}(\eta),g(\phi)\rangle] +
       E[\langle f_{\mvec}(\eta),g(\phi)\rangle ^* \Phi(f_{\nvec}(\eta))   ]   \Big)\nonumber\\
                && \phantom{XXXXXX}- \frac{f_{\mvec,r}^*(\eta)\cdot Tf_{\nvec,r}(\eta)}{\lambda^*_{\mvec}(\eta)+\lambda_{\nvec}(\eta)}, 
                                  \end{eqnarray}
the last term  in the third line being an Ito-term.  For this term $f_{\mvec,r}(\eta)$ and   $f_{\nvec,r}(\eta)$  are two-dimensional
vectors, and $T$ is to be interpreted as a $2\times2$-diagonal matrix, $T= {\rm Diag}(T_1,T_2)$.   For the linear theory,
the first term on the right side is absent.  The component $f _{\mvec,r}(\eta) = \eta\langle f_{\mvec,\pi}(\eta),\alpha\rangle/(\lambda_{\mvec}(\eta)+1)$
is of order ${\cal O} (\eta n^{\theta-1})$; a similar estimate holds for $f _{\mvec,r}(\eta)$.  We  therefore have that
\begin{equation}\label{stationident2.eq}
E[\Phi(f_{\mvec}(\eta))^*\Phi(f_{\nvec}(\eta))]= {\cal O}\left(\eta^2 \frac{m^{\theta-1} n^{\theta-1}}{m-n}\right).
\end{equation}
This equation, combined with  Eq.(\ref{matrixfar2.eq})  above, gives the estimate
\begin{eqnarray}
   \lefteqn{\langle e_{\mvec,\pi}(\eta),\partial_x e_{\nvec,\phi}(\eta)\rangle E[\Phi(f_{\mvec}(\eta))^*\Phi(f_{\nvec}(\eta))]= {\cal
      O}\left(\eta^4\frac{m^{2\theta-1}n^{2\theta-1}}{(m^2-n^2)(m-n)}\right)}\nonumber\\
         &&\phantom{XXXXXXXX}+{\cal
      O}\left(\eta^4\frac{m^{2\theta}n^{2\theta-2}}{(m^2-n^2)(m-n)}\right) + {\cal O}\left( \eta^6\frac{m^{3\theta -2} n^{3\theta-2}}{m-n}\right)    
\end{eqnarray}    
for $\mvec\neq \nvec$, at least in the situation where $m\neq n$.  This quantity is certainly  summable over $\mvec$ and $\nvec$, $m\neq n$ using
Young's inequality, the sum going to zero, $\eta\rightarrow 0$.  
                    
                    The ``antiresonant" terms in the double sum with $m=n$, but with different choices of the $\pm$ and no restriction on the $\sigma$'s, have matrix elements
$\langle e_{\mvec,\pi}(\eta),\partial_x e_{\nvec,\phi}(\eta)\rangle $ which can be ${\cal O}(1)$ in $n$.  But the covariance factors are ${\cal O}(\eta^2 n^{2\theta-3})$, from
the identity Eq.(\ref{stationident.eq}),
 with the denominator $\left(\lambda_{\mvec}^*(\eta) + \lambda_{\nvec}(\eta)\right)$ being ${\cal O}(n)$.   So again the products of the
matrix element and covariance factors for these terms are summable, the sum going to zero, $\eta\rightarrow 0$.  \\
                  
\noindent{\it Remark:} A term of the sort $E[\Phi(f_{\mvec}(\eta))^*\langle f_{\nvec}(\eta),g(\phi)\rangle]$ appearing in
the identity Eq.(\ref{stationident.eq}) is
no worse than $\kappa^2{\cal O}(n^{-1/2} )$, where $\kappa^2$ is a uniform bound on variances $E[|\Phi(f_{\nvec}(\eta))|^2 ]$ and $g$ is Lipschitz, a remark
potentially of utility in analyzing the non-linear problem.\\
 
    These off-diagonal terms in the double sum also include terms with matrix elements and field correlations
involving the eigenvectors $e_{-1,\sigma}(\eta)$ and
      $f_{-1,\sigma}(\eta)$,  corresponding to eigenvalues  near $-1$.
In terms of the $r$-components of $e_{-1,\sigma}(\eta)$ that are ${\cal O}(1)$ (actually nearly normalized),   the $\phi$-component is given by
\begin{equation}
e_{-1,\sigma,\phi}(\eta)= -\eta e_{-1,\sigma,r}(\eta)\cdot  (\lambda_{-1,\sigma}^2(\eta)-\partial_x^2 +{1})^{-1}\alpha.  
\end{equation}
The coefficient estimates Eq.(\ref{fouriercoefficients.eq}) then imply 
\begin{eqnarray}
   \langle
        e_{\nvec,\pi}(\eta),\partial_xe_{-1,\sigma,\phi}(\eta)\rangle &=&
        {\cal O}(\eta n^{\theta-1}) \nonumber\\ 
 \langle  e_{-1,\sigma,\pi}(\eta),\partial_xe_{\nvec,\phi}(\eta)\rangle
        &=&
        {\cal O}(\eta n^{\theta-2})\nonumber\\
 \langle e_{-1,\sigma,\pi}(\eta),\partial_xe_{-1,\sigma',\phi}(\eta)\rangle&=&
            {\cal O}(\eta^2)
\end{eqnarray}
and, from the identity Eq.(\ref{stationident.eq}),
\begin{eqnarray}
      E\left[\Phi(f_{\nvec}(\eta))^*\Phi(f_{-1,\sigma}(\eta))\right]&=&{\cal
      O}(\eta n^{\theta-1})
      \nonumber\\
      E\left[\Phi(f_{-1,\sigma}(\eta))^*\Phi(f_{-1,\sigma'}(\eta))\right]&=& {\cal O}(1).
      \end{eqnarray}
The resulting products such as  $ \langle
        e_{\nvec,\pi}(\eta),\partial_xe_{-1,\sigma,\phi}(\eta)\rangle
        E\left[\Phi(f_{\nvec}(\eta))^*\Phi(f_{-1,\sigma}(\eta))\right]$
        are clearly summable in $n$ and go to zero, $\eta\rightarrow0$, hence do
not contribute in the limit to the expected current.    \\

\noindent{\it Near-resonant terms}\\

Finally we investigate the near-resonant terms in the double sum of the sort
\begin{equation}\label{nearresonant.eq}
   \langle e_{\pm,n,1,\pi}(\eta), \partial_xe_{\pm,n,2,\phi}(\eta)\rangle E[\Phi(f_{\pm,n,1}(\eta))^*\Phi(f_{\pm,n,2}(\eta))].
   \end{equation}
We have by the Fourier coefficient estimates, Eq.(\ref{fouriercoefficients.eq}), that
\begin{eqnarray}\label{matrixnear.eq}
    \lefteqn {\langle e_{\pm,n,1,\pi}(\eta),\partial_xe_{\pm,n,2,\phi}(\eta)\rangle}\nonumber\\ 
       &=& \frac{n}{\lambda_{\pm,n,2}(\eta)}\Big(\langle e_{\pm,n,1,\pi}(0),e_{\pm,n,1,\pi}(0)\rangle + \langle e_{\pm,n,2,\pi}(0), (e_{\pm,n,2,\pi}(\eta)-e_{\pm,n,2,\pi}(0))\rangle \nonumber\\
&&  \phantom{X^t}+\langle e_{\pm,n,1,\pi}(\eta)-e_{\pm,n,1,\pi}(0), e_{\pm,n,1,\pi}(0)\rangle \Big)\nonumber\\
&& \phantom{X^t} +\frac{1}{\lambda_{\pm,n,2}(\eta)}\langle (e_{\pm,n,1,\pi}(\eta)-e_{\pm,n,1,\pi}(0)),\partial_x(e_{\pm,n,2,\pi}(\eta)-e_{\pm,n,2,\pi}(0))\rangle\nonumber\\
    &=&  \frac{n}{2\lambda_{\pm,n,2}(\eta)} + {\cal
    O}(\eta^4n^{4\theta-2}\ln^2n).
\end{eqnarray}    
We have used   $\partial_xe_{\pm,n,2,\pi}(0)
= n e_{\pm,n,1,\pi}(0)$ and $\partial_xe_{\pm,n,1,\pi}(0)
= -n e_{\pm,n,2,\pi}(0)$. Recall as well that the $e_{\nvec}(0)$'s are
normalized so that  $\langle e_{\pm,n,1,\pi}(0),e_{\pm,n,1,\pi}(0)\rangle =
1/2$, and that $\lambda_{\nvec}(\eta)= \lambda_{\nvec}(0)+{\cal
  O}(\eta^2 n^{2\theta-1})$ with $\lambda_{\nvec}(0)=\pm i(n^2+1)^{1/2}$. We summarize:
\begin{equation}\label{nearresmatrixel2.eq}
 \langle e_{\pm,n,1,\pi}(\eta),\partial_xe_{\pm,n,2,\phi}(\eta)\rangle=\pm 
         \frac{n}{2i(n^2+1)^{1/2}}+{\cal O}(\eta^2
 n^{2\theta-2})+{\cal O}(\eta^4 n^{4\theta-2}\ln^2n).
\end{equation}

It remains to analyze the  covariance factor  $E[\Phi(f_{\pm,n,1}(\eta)^*\Phi(f_{\pm,n,2}(\eta))]$ in Eq.(\ref{nearresonant.eq}).  
Again, we assume that there is no non-linearity in the identity Eq.(\ref{stationident.eq}) so that
\begin{eqnarray}\label{neardiagff.eq}
         \lefteqn{E[\Phi(f_{\pm,n,1}(\eta))^*\Phi(f_{\pm,n,2}(\eta))] = -\frac{f^*_{\pm,n,1,r}(\eta)\cdot T f_{\pm,n,2,r}(\eta) }{\lambda^*_{\pm,n,1}(\eta)+\lambda_{\pm,n,2}(\eta)} } \nonumber\\
         &=&-\frac{\eta^2 \langle f_{\pm,n,1,\pi}(\eta),\alpha\rangle^*\cdot T \langle f_{\pm,n,2,\pi}(\eta),\alpha\rangle}{(\lambda^*_{\pm,n,1}(\eta)+\lambda_{\pm,n,2}(\eta))(\lambda^*_{\pm,n,1}(\eta)+1)(\lambda_{\pm,n,2})(\eta)+1)}
\end{eqnarray}
 ($\langle f_{\nvec,\pi}(\eta),\alpha_i\rangle$ is regarded as a two-component vector, $i=1,2$).  

The numerator factor on the right side of this expression
 can be expanded:
 \begin{eqnarray}\label{TIto.eq}
         \langle f_{\pm,n,1,\pi}(\eta),\alpha\rangle^*\cdot T \langle f_{\pm,n,2,\pi}(\eta),\alpha\rangle&=&\frac{(T_1-T_2)}{2} \langle f_{\pm,n,1,\pi}(\eta),\alpha\rangle^*\cdot D \langle f_{\pm,n,2,\pi}(\eta),\alpha\rangle\nonumber\\
         &&\!\!\!\!\!\!\!\!\!\!+\frac{(T_1+T_2)}{2}\langle f_{\pm,n,1,\pi}(\eta),\alpha\rangle^*\cdot  \langle f_{\pm,n,2,\pi}(\eta),\alpha\rangle,
  \end{eqnarray}
with $D$ the two-dimensional diagonal matrix $\rm{Diag}(1,-1)$.  But again, using the Fourier coefficient estimates Eq.(\ref{fouriercoefficients.eq}), which are
also satisfied by the momentum components $\{f_{\nvec,\pi}(\eta)\}$ of the $f_{\nvec}$'s, we find this equal to
 \begin{equation}\label{TIto2.eq}
     \frac{(T_1-T_2)}{2} \langle f_{\pm,n,1,\pi}(0),\alpha\rangle^*\cdot D \langle f_{\pm,n,2,\pi}(0),\alpha\rangle 
         +{\cal O}(\eta^2 n^{4\theta-1}\ln n).
  \end{equation}
Here, $ \langle f_{\pm,n,1,\pi}(\eta),\alpha\rangle$ and $\langle f_{\pm,n,2,\pi}(\eta),\alpha\rangle$ have nearly zero dot product (exactly zero for $\eta= 0$), 
so that the  $(T_1+T_2)$ part is represented in the small remainder terms.  Moreover, via two-dimensional matrix algebra, 
\begin{equation}\label{Dcal.eq}
\nu_n\equiv \langle f_{\pm,n,1,\pi}(0),\alpha\rangle^*\cdot D \langle
   f_{\pm,n,2,\pi}(0),\alpha\rangle= 
\frac{\Im\left(\hat{\alpha}_1^{*2}(n)
   \hat{\alpha}_2^{2}(n)\right)}{\left|\hat{\alpha}_1^2(n)
+ \hat{\alpha}_2^2(n)\right|}.
\end{equation}
(The notation $v_n$ is temporarily introduced to facilitate handling of already complicated expressions.)
Thus the numerator factor is given by
\begin{eqnarray}\label{TIto3.eq}
         \lefteqn{\langle f_{\pm,n,1,\pi}(\eta),\alpha\rangle^*\cdot T \langle f_{\pm,n,2,\pi}(\eta),\alpha\rangle}\nonumber\\
         &=&\!\!\frac{(T_1-T_2)}{2}  \nu_n        +{\cal O}(\eta^2 n^{2\theta-1}\ln n).
  \end{eqnarray}
  
 The denominator of Eq.(\ref{neardiagff.eq}) is expanded with the aid of the eigenvalue expansion Eq.(\ref{eigenshift5appendix.eqn10}):
\begin{eqnarray}
\label{bigfrac.eq}
        \lefteqn{\frac{\eta^2}{(\lambda_{(\pm,n,1)}(\eta)^*+\lambda_{\pm,n,2}(\eta))(\lambda_{\pm,n,1}(\eta)^*+1)(\lambda_{\pm,n,2}(\eta)+1)}}\nonumber\\
&=& \frac{-2}{(\mu_{n,1}+\mu_{n,2})\pm
	  i(n^2+1)^{1/2}(\mu_{n,1}-\mu_{n,2})} 
             +{\cal O}(\eta^2 n^{-2}\ln n).
\end{eqnarray}

Multiplying the factors  (\ref {TIto3.eq}) and (\ref{bigfrac.eq}) in (\ref{neardiagff.eq}) and noting that $\nu_n= {\cal O}(n^{2\theta})$,   we obtain
\begin{eqnarray}\label{neardiagff3.eq}
        \lefteqn{E[\Phi(f_{\pm,n,1}(\eta))^*\Phi(f_{\pm,n,2}(\eta))]}&&\nonumber\\  
   &=&\frac{(T_1-T_2)}{2}\frac{2\nu_n }{ \left((\mu_{n,1}+\mu_{n,2})\pm
	  i(n^2+1)^{1/2}(\mu_{n,1}-\mu_{n,2})\right) }\nonumber\\
	&& +{\cal O}(\eta^2n^{-2}\ln n ).
	      \end{eqnarray}
  
Equation  (\ref{nearresmatrixel2.eq})  and the previous Eq.(\ref{neardiagff3.eq}) show that
\begin{eqnarray}\label{nearresonant2.eq}
  \lefteqn{\langle e_{\pm,n,1,\pi}(\eta),
   \partial_xe_{\pm,n,2,\phi}(\eta)\rangle
   E[\Phi(f_{\pm,n,1}(\eta))^*\Phi(f_{\pm,n,2}(\eta))]}\nonumber\\ 
   &=&-\frac{T_1-T_2}{2} \frac{n}{(n^2+1)^{1/2}}\frac{\nu_n}{\left(
	  (n^2+1)^{1/2}(\mu_{n,1}-\mu_{n,2})\mp i(\mu_{n,1}+\mu_{n,2})\right)}      \nonumber\\
	&&\phantom{XXXXX}  +{\cal O} (\eta^2n^{-2}\ln n ) +{\cal O}(\eta^2 n^{2\theta-3}) +{\cal O}(\eta^4 n^{4\theta-3}\ln^2n).
\end{eqnarray}

\noindent{\it Symmetries}\\
 
In the limit, $\eta\rightarrow 0$ the following
symmetries obtain:
\begin{eqnarray}\label{symm.eq}
\lefteqn{\langle e_{\pm,n,1,\pi}(0),
   \partial_xe_{\pm,n,2,\phi}(0)\rangle
   E[\Phi(f_{\pm,n,1}(0))^*\Phi(f_{\pm,n,2}(0))]}&&\nonumber\\ 
&=&\Big(\langle e_{\pm,n,2,\pi}(0),
   \partial_xe_{\pm,n,1,\phi}(0)\rangle
   E[\Phi(f_{\pm,n,2}(0))^*\Phi(f_{\pm,n,1}(0))]\Big)^*\nonumber\\
 &=&\Big(\langle e_{\mp,n,2,\pi}(0),
   \partial_xe_{\mp,n,1,\phi}(0)\rangle
   E[\Phi(f_{\mp,n,2}(0))^*\Phi(f_{\mp,n,1}(0))]\Big),   
 \end{eqnarray}            
the first relation involving an integration by parts in the matrix element and using that  $e_{\nvec,\phi}(0)= \frac{1}{\lambda_{\nvec}(0)}e_{\nvec,\pi}(0)$.
 The third line follows from the facts that
$e_{\pm,n,\sigma,\phi}(0)= -e_{\mp,n,\sigma,\phi}(0)$ and an integration by parts so that
\begin{equation}
\langle e_{\mp,n,2,\pi}(0),
   \partial_xe_{\mp,n,1,\phi}(0)\rangle= \langle e_{\pm,n,1,\pi}(0),
   \partial_xe_{\pm,n,2,\phi}(0)\rangle,
   \end{equation}
    and 
   \begin{equation}
  E[ \Phi(f_{\mp,n,2}(0))^*\Phi(f_{\mp,n,1}(0))] = E[ \Phi(f_{\pm,n,1}(0))^*\Phi(f_{\pm,n,2}(0))],
\end{equation}
 from Eq.(\ref{neardiagff.eq}).
This simplifies the residual part of the double sum in Eq.(\ref{doublesum.eqn1}); the expected current is, in any case, real.

In summary, the  diagonal 
and off-diagonal terms and the
remainder part of the near resonant
terms in (\ref{nearresonant2.eq})
in the double sum Eq.(\ref{doublesum.eqn1}) are uniformly summable,
and these sums go to zero, $\eta \rightarrow 0$. With the symmetry
of the surviving terms, we have the final result 
\begin{eqnarray}
\label{doublesum.eqn2}
  \lefteqn{\lim_{\eta\rightarrow0} E\left[\frac{1}{2\pi}\int_0^{2\pi}\!\!\!\!\!\!\pi(x)\partial_x\phi(x)\,dx\right]}&&\nonumber\\
&=&
 \frac{1}{2\pi} \lim_{\eta\rightarrow0}\sum_{\mvec}\sum_{\nvec}\langle e_{\mvec,\pi}(\eta),\partial_xe_{\nvec,\phi}(\eta)\rangle
  E[\Phi(f_{\mvec}(\eta))^*\Phi(f_{\nvec}(\eta))] \nonumber\\
&=&-\frac{(T_1-T_2)}{2}\frac{1}{2\pi}\sum_{n:\,n>0} \frac{4n\nu_n(\mu_{n,1}-\mu_{n,2})}{\left(
	  (n^2+1)(\mu_{n,1}-\mu_{n,2})^2
    +(\mu_{n,1}+\mu_{n,2})^2\right)}\nonumber\\
&\equiv&-\frac{(T_1-T_2)}{2}\frac{1}{\pi}\sum_{n:\,n>0} \frac{n \Im\left(\hat{\alpha}_1^{*\,2}(n) \hat{\alpha}_2^2(n) \right)}
 {(n^2+1)|\hat{\alpha}_1^2(n) +\hat{\alpha}_2^2(n)|^2 +(|\hat{\alpha}_1^2(n)| +|\hat{\alpha}_2^2(n)|)^2}
\end{eqnarray}
by the definition of   the $\mu_{n,\sigma}$'s and $\nu_n$.  This concludes the proof of the theorem. \qed

\acknowledgements The author wishes to acknowledge discussions with
Luc Rey-Bellet  and to acknowledge particularly the prior work with Yao Wang, with whom the current was first found.

\end{document}